\documentclass[a4paper]{article}
\usepackage{INTERSPEECH2021}

\usepackage{cite}
\usepackage{amsmath,amssymb,amsfonts}
\usepackage{algorithmic}
\usepackage{graphicx}
\usepackage{textcomp}
\usepackage{xcolor}
\usepackage{url}
\usepackage{booktabs}
\usepackage[acronym, nomain, nopostdot]{glossaries}
\usepackage{color}
\usepackage{import}
\usepackage{tikz, pgfplots, pgfkeys, pstricks}
\usetikzlibrary{arrows.meta}
\pgfplotsset{compat=1.16}

\newacronym{ADRENALINE}{ADRENALINE}{\emph{\underline{A}ttention-based \underline{D}eep \underline{RE}current \underline{N}eural-network for loc\underline{ALI}zing sou\underline{N}d \underline{E}vents}}
\newacronym{CNN}{CNN}{convolutional neural network}
\newacronym{DoA}{DoA}{direction-of-arrival}
\newacronym{GMM}{GMM}{Gaussian mixture model}
\newacronym{GRU}{GRU}{gated recurrent unit}
\newacronym{KF}{KF}{Kalman filter}
\newacronym{KLD}{KLD}{Kullback-Leibler divergence}
\newacronym{LGS}{LGS}{linear Gaussian system}
\newacronym{LSTM}{LSTM}{long-short term memory}
\newacronym{MSE}{MSE}{mean squared error}
\newacronym{MUSIC}{MUSIC}{multiple signal classification}
\newacronym{NLP}{NLP}{natural language processing}
\newacronym{PILOT}{PILOT}{\emph{\underline{P}}robab\emph{\underline{i}}listic \emph{\underline{L}}ocalization of S\emph{\underline{o}}unds with \emph{\underline{T}}ransformers}
\newacronym{ReLU}{ReLU}{rectified linear unit}
\newacronym{RNN}{RNN}{recurrent neural network}
\newacronym{SEL}{SEL}{sound event localization}
\newacronym{SELD}{SELD}{sound event localization and detection}
\newacronym{SRP-PHAT}{SRP-PHAT}{steered  response  power  phase  transform}
\newacronym{STFT}{STFT}{short-time Fourier transform}

\title{PILOT: Introducing Transformers for Probabilistic Sound Event Localization}
\name{Christopher Schymura$^1$, 
Benedikt Bönninghoff~$^1$, 
Tsubasa Ochiai$^2$,
Marc Delcroix$^2$, \\
Keisuke Kinoshita$^2$,
Tomohiro Nakatani$^2$,
Shoko Araki$^2$,
Dorothea Kolossa$^1$
}
\address{
  $^1$Cognitive Signal Processing Group, Ruhr University Bochum, Bochum, Germany\\
  $^2$NTT Communications Science Laboratories, NTT Corporation, Kyoto, Japan}
\email{\{christopher.schymura, benedikt.boenninghoff, dorothea.kolossa\}@rub.de
\\
\{tsubasa.ochiai.ah, keisuke.kinoshita.mb, shoko.araki.pu\}@hco.ntt.co.jp 
\\
\{marc.delcroix, tnak\}@ieee.org
}
\begin{document}

\maketitle
\begin{abstract}
Sound event localization aims at estimating the positions of sound sources in the environment with respect to an acoustic receiver (e.g. a microphone array). Recent advances in this domain most prominently focused on utilizing deep recurrent neural networks. Inspired by the success of transformer architectures as a suitable alternative to classical recurrent neural networks, this paper introduces a novel transformer-based sound event localization framework, where temporal dependencies in the received multi-channel audio signals are captured via self-attention mechanisms. Additionally, the estimated sound event positions are represented as multivariate Gaussian variables, yielding an additional notion of uncertainty, which many previously proposed deep learning-based systems designed for this application do not provide. The framework is evaluated on three publicly available multi-source sound event localization datasets and compared against state-of-the-art methods in terms of localization error and event detection accuracy. It outperforms all competing systems on all datasets with statistical significant differences in performance.
\end{abstract}
\noindent\textbf{Index Terms}: Audio signal processing, self attention, probabilistic deep learning, sound event localization, transformer

\section{Introduction}
\label{sec:introduction}
The ability of a computational system to locate sounds in the environment, often termed \gls{SEL}, enables a variety of technical applications. In many cases, a sound event refers to a speech source, which is important in applications such as speech enhancement~\cite{Li2018, Liu2019}, speaker and meeting diarization~\cite{Araki2008, Schmalenstroeer2010}, automatic speech recognition~\cite{Woelfel2006} and many others. However, a sound event can also be associated with a broader spectrum of acoustic sources, enabling the application of \gls{SEL} in domains like industrial monitoring~\cite{Grobler2017} and environmental traffic analysis~\cite{Kodera2007}. Specific challenges of \gls{SEL} include acoustic disturbances caused by background noise and reverberation, as well as the potential temporal overlap of simultaneously active acoustic sources.

\begin{figure}[t]
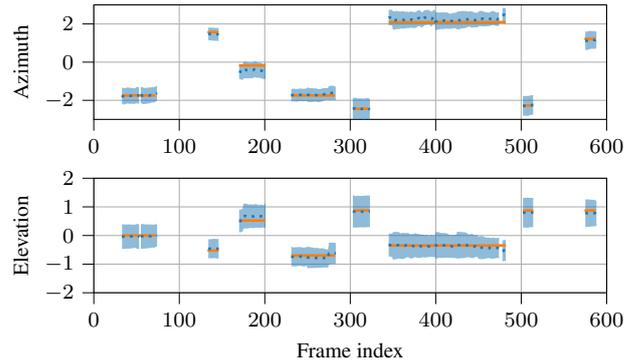

	\centering
	\begin{minipage}{\textwidth}
		\input{figs/blaster_azimuth}%
	\end{minipage}
	\begin{minipage}{\textwidth}
		\input{figs/blaster_elevation}%
	\end{minipage}
	\caption{Azimuth and elevation estimates (blue dotted lines) in radians, provided by the proposed PILOT framework. The solid orange lines show ground-truth sound event \glspl{DoA}. The estimates provided by PILOT are associated with uncertainty measures of the underlying Gaussian model, depicted as shaded blue regions.\vspace{-0.5cm}}
	\label{fig:estimation_comparison}
\end{figure}

These challenges have been addressed by classical \gls{SEL} systems primarily based on parametric approaches~\cite{Schmidt1986, DiBiase2001, Sawada2005}, as well as by more recently introduced deep learning frameworks~\cite{Chakrabarty2017, He2018, Adavanne2018b, Adavanne2018a}. Regardless of the specific architectural considerations, many \gls{SEL} systems incorporate temporal context into the estimation process. While classical methods usually rely on recursive Bayesian estimation techniques~\cite{Klee2006, Hao2014}, modern systems make use of recurrent structures like \gls{LSTM} networks~\cite{Hochreiter1997}. However, conventional neural network-based systems lack the ability to associate uncertainty with the estimated sound event locations; a feature that traditional Bayesian approaches naturally provide. Additionally, \glspl{RNN} are often struggle to learn long-term dependencies within time-series data~\cite{Trinh2018}.

This paper proposes a novel deep learning framework for \gls{SEL} that adopts a transformer-based architecture~\cite{Vaswani2017} with a probabilistic representation of sound event locations. Therefore, it will be abbreviated as \gls{PILOT}. Following the design philosophy of the transformer, the proposed model provides a conventional feed-forward signal flow and handles temporal dependencies via self-attention mechanisms. The probabilistic output is based on a differentiable \gls{LGS}~\cite[Chap.~4]{Murphy2013} model that is adopted in this work to handle multi-source \gls{SEL} scenarios. Fig.~\ref{fig:estimation_comparison} depicts exemplary trajectories of \gls{DoA} estimates provided by \gls{PILOT}, which illustrate the models capability of associating uncertainty measures with the corresponding predictions.

\begin{figure*}[t]
	\centering
	\includegraphics[width=16.5cm]{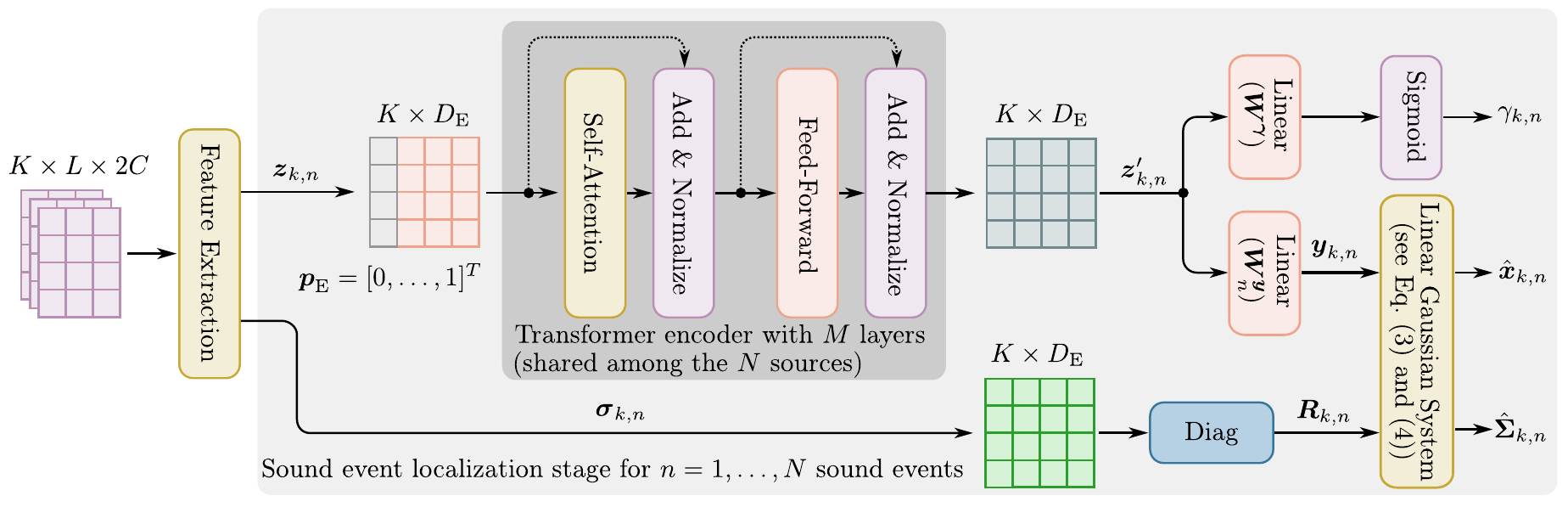}
	\vspace*{-0.2cm}
	\caption{Overview of the PILOT model. An input tensor of the \gls{STFT} representation with \(K\) time frames, \(L\) frequency bins and \(2C\) channels (containing both magnitude and phase components), is fed into a \gls{CNN}-based feature extraction stage. Positional encodings are attached to the extracted features, resulting in a \(D_{\mathrm{E}}\)-dimensional intermediate representation, which serves as input to the transformer network. The output is then further processed, yielding indicator variables on sound event activity \(\gamma_{k, n}\) and their corresponding locations. The latter are represented as multivariate Gaussians with mean~\(\hat{\boldsymbol{x}}_{k, n}\) and covariance matrix \(\hat{\boldsymbol{\Sigma}}_{k, n}\) for each of the \(n=1,\,\ldots,\,N\) sound events that can be assessed by the model.\vspace{-0.3cm}}
	\label{fig:system_overview}
\end{figure*}%

\section{Related work}
\label{sec:related_work}
Deep learning-based systems for \gls{SEL} most prominently rely on \glspl{CNN}, \glspl{RNN} or a combination of both. For instance, the model proposed in~\cite{Chakrabarty2017} uses a \gls{CNN}-based feature extraction, which operates on the \gls{STFT} spectrum to predict the \gls{DoA} of a sound event. A similar approach was introduced in~\cite{Adavanne2018b}, where the \gls{CNN} uses the \gls{STFT} representation to generate a spatial pseudo spectrum, resembling the \gls{DoA} spectrum obtained by the \gls{MUSIC} algorithm. An extension towards handling \gls{SELD} was proposed in~\cite{Adavanne2018a}. Herein, the task focuses on estimating the association of the sound event type, jointly with its position. The proposed system solved this task via a combination of \gls{CNN}-based feature extraction and \gls{RNN}-based estimation of spatial locations and classification of event types. A related system for \gls{SEL} was proposed in~\cite{Schymura2020}, which adopted the \gls{CNN}-based feature extraction, but exchanged the \gls{RNN} stage with an attention-based sequence-to-sequence architecture~\cite{Cho2014b}. Additionally, approaches to joining deep learning with classical probabilistic models have recently gained attention. Notable examples are the backprop \gls{KF}~\cite{Haarnoja2016}, the differentiable particle filter~\cite{Jonschkowski2018} and recurrent Kalman networks~\cite{Becker2019}. Related to the task of \gls{SEL}, a \gls{KF}-based deep learning framework was proposed in~\cite{Schymura2020b} for audiovisual speaker tracking. Other probabilistic models to tackle \gls{SEL} are described in, e.g.,~\cite{Deleforge2015, Braun2019}.

\section{System description}
\label{sec:system_description}
The framework proposed in this work is composed of three individual stages: a \gls{CNN}-based feature extraction stage, a transformer-based encoder to handle temporal dependencies in the audio signals and a probabilistic output stage that indicates active sound events and their respective spatial locations. An overview of the complete system is depicted in Fig.~\ref{fig:system_overview}.

\subsection{Feature extraction}
\label{subsec:feature_extraction}
The \gls{CNN}-based feature extractor from SELDNet~\cite{Adavanne2018a} is utilized in this work. An acoustic waveform with \(C = 4\) channels is divided into non-overlapping chunks of \(500\,\text{ms}\) duration. An \gls{STFT} with a 2048-point Hamming window, a frame length of \(40\,\text{ms}\) and \(20\,\text{ms}\) shift, is performed on each chunk. Magnitude and phase of the complex spectrum are concatenated to an input tensor of dimension \(K \times L \times 2C\), where \(K = 25\) is the number of frames within one chunk and \(L = 1024\) is the number of non-redundant frequency bins.

A three-layer \gls{CNN} is employed to derive features from the \gls{STFT} representation. Each layer contains \(64\) filters with kernel size \(3 \times 3\), followed by batch normalization, \glspl{ReLU} and max-pooling. The latter has a kernel size of \(1 \times 8\) in the first two layers and \(1 \times 2\) in the third layer. The third layer output is flattened and fed into a fully-connected network with three layers, intermediate hidden dimensions of 128 and \glspl{ReLU} as hidden activations. The output layer of this network is composed of two segments: The first segment with linear activations returns feature vectors \(\boldsymbol{z}_{k, n} \in \mathbb{R}^{D_{\mathrm{F}}}\). The second segment produces estimates of the observation noise variances \(\boldsymbol{\sigma}_{k, n} \in \mathbb{R}^{D_{\mathrm{E}}}\) with \(D_{\mathrm{E}} = D_{\mathrm{F}} + 1\), using exponential non-linearities. Both segments produce these outputs for each of the \(n = 1,\,\ldots,\,N\) sound events, where $N$ denotes the maximum number of sound events that the system can handle simultaneously.

\subsection{Transformer network for handling temporal context}
\label{subsec:transformer}
A transformer encoder 
as introduced in~\cite{Vaswani2017} is the primary building block of the intermediate stage. Each sequence of feature vectors \(\{\boldsymbol{z}_{k, n}\}_{k=1}^{K}\) is concatenated with a positional encoding vector \(\boldsymbol{p}_{\mathrm{E}} = \begin{bmatrix}p_{\mathrm{E}, 1} & \cdots & p_{\mathrm{E}, K}\end{bmatrix}^{\mathrm{T}} \in \mathbb{R}^{K}\) with elements \(p_{\mathrm{E}, k} = \frac{1}{K-1} (k-1)\), yielding a \(D_{\mathrm{E}}\) dimensional tensor, as depicted in Fig.~\ref{fig:system_overview}. This approach to encode information about the temporal order differs from the positional encodings described in~\cite{Vaswani2017}. Initial experiments have shown that a linearly increasing vector resulted in superior performance over positional encodings based on trigonometric functions.

The transformer encoder itself follows a conventional architecture with \(M = 3\) layers. Each layer incorporates a self-attention network with \(4\) heads, followed by layer normalization, a fully-connected feed-forward network with an intermediate dimension of 1024 and another layer normalization stage at the output. Both self-attention and feed-forward networks can be bypassed by additional skip connections, exactly following the specifications outlined in~\cite{Vaswani2017}. The transformer output has the same dimensions as the input and will be denoted as \(\{\boldsymbol{z}_{k, n}^{\prime}\}_{k=1}^{K}\), serving as input to the subsequent localization output stage of the \gls{PILOT} model.

\subsection{Probabilistic localization output stage}
\label{subsec:localization_output}
This stage is based on the ADRENALINE system proposed in~\cite{Schymura2020}. It is composed of two components: a vector \(\boldsymbol{\gamma}_{k} \in \mathbb{R}^{N}\) with elements \(\gamma_{k, n} \in [0,\,1]\), indicating sound event activity, where~\(N\) is the maximum number of sound events that the system can handle simultaneously. This vector is obtained as \(\boldsymbol{\gamma}_{k} = \sigma( \boldsymbol{W}^{\boldsymbol{\gamma}}\boldsymbol{z}_{k}^{\prime})\), where \(\boldsymbol{W}^{\boldsymbol{\gamma}} \in \mathbb{R}^{N \times D_{\mathrm{E}}}\) is a weight matrix and~\(\sigma(\cdot)\) denotes the sigmoid function. The spatial position of each sound event is encoded as a multivariate Gaussian variable with mean \(\hat{\boldsymbol{x}}_{k, n}\) and covariance matrix \(\hat{\boldsymbol{\Sigma}}_{k, n}\). The spatial coordinates are expressed via a \gls{DoA} representation \(\hat{\boldsymbol{x}}_{k, n} = \begin{bmatrix}
\hat{\phi}_{k, n} & \hat{\vartheta}_{k, n}\end{bmatrix}^{\mathrm{T}}\) with azimuth and elevation angles~\(\hat{\phi}_{k, n}\) and~\(\hat{\vartheta}_{k, n}\), respectively. 

A prominent advantage of the transformer-based model compared to \glspl{RNN} is its handling of temporal dependencies in sequence data without recurrent structures. Therefore, using recursive Bayesian estimation techniques in the output stage would render this advantage void. Hence, a \gls{LGS} is used here, which encodes the individual sound event \glspl{DoA} as
\begin{align}
p(\boldsymbol{x}_{k, n}) &= \mathcal{N}\Big(\boldsymbol{x}_{k, n}\,|\,\boldsymbol{\mu}_{n},\,\boldsymbol{\Sigma}_{n}\Big), \label{eqn:state_equation} \\
p(\boldsymbol{y}_{k, n}\,|\,\boldsymbol{x}_{k, n}) &= \mathcal{N} \Big(\boldsymbol{y}_{k, n}\,|\,\boldsymbol{C} \boldsymbol{x}_{k, n} + \boldsymbol{b},\,\boldsymbol{R}_{k, n}\Big),
\label{eqn:observation_equation}
\end{align}
\noindent where \(\boldsymbol{x}_{k, n} = \begin{bmatrix}
\phi_{k, n} & \vartheta_{k, n}\end{bmatrix}^{\mathrm{T}}\) is the \gls{DoA} vector corresponding to the \(n\)-th sound event. A multivariate Gaussian prior with mean~\(\boldsymbol{\mu}_{n} \in \mathbb{R}^{2}\) and covariance matrix~\(\boldsymbol{\Sigma}_{n} \in \mathbb{R}^{2 \times 2}\) is imposed on each \gls{DoA} vector in Eq.~\eqref{eqn:state_equation}. Eq.~\eqref{eqn:observation_equation} of the \gls{LGS} represents a Gaussian observation model, parameterized by an observation matrix~\(\boldsymbol{C} \in \mathbb{R}^{D_{\mathrm{E}} \times 2}\), bias term~\(\boldsymbol{b} \in \mathbb{R}^{D_{\mathrm{E}}}\) and observation noise covariance matrix~\(\boldsymbol{R}_{k, n} = \mathrm{diag}(\boldsymbol{\sigma}_{k, n})\). The observation vector~\(\boldsymbol{y}_{k, n}\) for the \(n\)-th source is obtained via a linear transformation \(\boldsymbol{y}_{k, n} = \boldsymbol{W}_{n}^{\boldsymbol{y}}\boldsymbol{z}_{k, n}^{\prime}\) with weights \(\boldsymbol{W}_{n}^{\boldsymbol{y}} \in \mathbb{R}^{D_{\mathrm{E}} \times D_{\mathrm{E}}}\). 

An inference scheme to obtain the posterior distribution \(p(\boldsymbol{x}_{k, n}\,|\,\boldsymbol{y}_{k, n}) \propto p(\boldsymbol{y}_{k, n}\,|\,\boldsymbol{x}_{k, n})p(\boldsymbol{x}_{k, n})\) can be derived by computing the first and second derivatives of the log-posterior using Eqs.~\eqref{eqn:state_equation} and \eqref{eqn:observation_equation}. Setting the first derivative to zero and solving for \(\hat{\boldsymbol{x}}_{k, n}\) yields the posterior mean
\begin{equation}
\hat{\boldsymbol{x}}_{k, n} = \hat{\boldsymbol{\Sigma}}_{k, n}\Big(\boldsymbol{\Sigma}_{n}^{-1}\boldsymbol{\mu}_{n} + \boldsymbol{C}^{\mathrm{T}} \boldsymbol{R}_{k, n}^{-1} (\boldsymbol{y}_{k, n} - \boldsymbol{b})\Big). \label{eqn:inference_mean}
\end{equation}
\noindent The second derivative is the curvature of the quadratic function describing the posterior distribution~\cite[Chap.~3]{Thrun2005}, whose inverse is the posterior covariance matrix
\begin{equation}
\hat{\boldsymbol{\Sigma}}_{k, n} = \Big(\boldsymbol{\Sigma}_{n}^{-1} + \boldsymbol{C}^{\mathrm{T}} \boldsymbol{R}_{k, n}^{-1} \boldsymbol{C}\Big)^{-1}. \label{eqn:inference_covariance}
\end{equation}
\noindent The inference scheme is fully differentiable, which makes it possible to directly utilize Eqs.~\eqref{eqn:inference_mean} and~\eqref{eqn:inference_covariance} for learning the previously described model parameters via back-propagation.

\subsection{Loss function}
\label{subsec:loss_function}
A modified version of the \gls{SEL} loss function introduced in~\cite{Schymura2020} is utilized in this work, which additionally takes into account the probabilistic nature of the localization output stage. The probabilistic \gls{SEL} loss is computed as
\begin{equation}
\mathcal{L}_{n} = \mathcal{L}_{\text{ACT}}(\hat{\gamma}_{n},\gamma_{n}) + \alpha \mathcal{L}_{\text{DOA}}(\hat{\boldsymbol{x}}_{n},\boldsymbol{x}_{n},\gamma_{n}) + \beta \mathcal{L}_{\mathrm{KL}}(\hat{\boldsymbol{\Sigma}}_{n})
\label{eqn:loss_function}
\end{equation}
for each of the~\(n=1,\,\ldots,\,N\) sound sources that can be represented by the model. The discrete time index \(k\) is omitted here and in the following for notational convenience and \(\alpha,\,\beta \in \mathbb{R}_{+}\) are scaling factors. Eq.~\eqref{eqn:loss_function} incorporates a binary cross-entropy loss term \(\mathcal{L}_{\text{ACT}}(\hat{\gamma}_{n},\gamma_{n})\) between estimated and ground-truth source activities~\(\hat{\gamma}_{n}\) and~\(\gamma_{n}\), as well as a \gls{DoA} loss term \(\mathcal{L}_{\text{DOA}}(\hat{\boldsymbol{x}}_{n},\boldsymbol{x}_{n},\gamma_{n}) = \xi_{n} \gamma_{n}\) utilizing the \gls{DoA} error 
\begin{equation}
\xi_{n} = \mathrm{acos}(\sin(\hat{\phi}_{n}) \sin(\phi_{n}) + \cos(\hat{\phi}_{n}) \cos(\phi_{n}) \cos(\Delta\vartheta_{n})),
\label{eqn:doa_error}
\end{equation}
with \(\Delta\vartheta_{n} = \vartheta_{n} - \hat{\vartheta}_{n}\). The term \(\mathcal{L}_{\mathrm{KL}}(\hat{\boldsymbol{\Sigma}}_{n})\) is required to prevent the system from setting the parameterized covariance matrices to zero. It models the \gls{KLD} between \(p(\boldsymbol{x}_{k, n}\,|\,\boldsymbol{y}_{k, n})\) and a corresponding multivariate Gaussian distribution with identical mean and unit covariance matrix, similar to the penalty term utilized for training variational autoencoders~\cite{Kingma2014b}. The final loss is obtained by summing the individual losses in Eq.~\eqref{eqn:loss_function} over all time steps and sources, utilizing permutation invariant training \cite{7952154}.

\section{Evaluation}
\label{sec:evaluation}
The proposed \gls{PILOT} framework is evaluated on three publicly available datasets. The detailed training procedure and experimental setup is described in the following. The program code is accessible online\footnote{\url{https://github.com/rub-ksv/pilot}}.

\subsection{Datasets}
\label{subsec:datasets}
The ANSYN, RESYN and REAL datasets presented in~\cite{Adavanne2018a} are used as evaluation corpora in this work. All three datasets are composed of multi-channel audio signals in first-order Ambisonics format. The first two datasets use simulated anechoic and reverberant impulse responses, whereas the latter one is based on impulse responses recorded in realistic environments. All datasets comprise three subsets each, corresponding to either no, at most two, or at most three temporally overlapping sources. Each subset provides three cross-validation splits with 300 audio files. These files are divided into 240 files for training and 60 for validation. The audio signals are sampled with \(44.1\,\text{kHz}\) and have a duration of \(30\,\text{s}\) each. The audio signals in the synthetic datasets contain sound events from 11 different classes, covering the full azimuth range and an elevation range from \(-60^{\circ}\) to \(60^{\circ}\). The REAL dataset utilizes sound events from 8 classes. It covers the full azimuth range and the elevation range was restricted between \(-40^{\circ}\) and \(40^{\circ}\).

\subsection{Baseline methods}
\label{subsec:baseline_methods}
Three baseline methods were employed in this work. The \gls{CNN} feature extraction stage described in Sec.~\ref{subsec:feature_extraction} is used as a first baseline to establish a lower bound on performance. Herein, all intermediate steps between the feature extraction and the output stage are omitted, yielding a small model with 110668 parameters. Additionally, the modified SELDNet architecture~\cite{Adavanne2018a} (without event type association to guarantee a fair comparison) and the ADRENALINE framework described in~\cite{Schymura2020} constitute the second and third baseline models with 551500 and 448204 parameters, respectively. Both architectures employ recurrent structures via \glspl{GRU} and the latter method additionally makes use of attention mechanisms. The proposed \gls{PILOT} framework incorporates 466736 trainable parameters in total.

\subsection{Performance metrics}
\label{subsec:metrics}
Frame recall and \gls{DoA} error are used as performance metrics. The frame recall metric~\cite{Adavanne2018a} indicates the percentage of frames, where the number of active sources was estimated correctly. A source is considered active if its indicator variable in the source activity vector exceeds a threshold of \(0.5\). The \gls{DoA} error on the test set is computed using the definition in Eq.~\eqref{eqn:doa_error}. Herein, all active sources at a particular frame are incorporated and the Hungarian algorithm~\cite{Kuhn1955} is used to solve the assignment problem between multiple estimated \glspl{DoA} and the target \glspl{DoA}.

\subsection{Experimental setup}
\label{subsec:experimental_setup}
Both CNN and SELDNet baselines were parameterized as reported in~\cite{Adavanne2018a}. The parameters of the ADRENALINE baseline model were also specified as outlined in the corresponding publication~\cite{Schymura2020}. The loss term in Eq.~\eqref{eqn:loss_function} uses internal scaling factors of \(\alpha = \beta = 1\). The Gaussian prior in Eq.~\eqref{eqn:state_equation} was initialized with equally spaced azimuth angles and elevation set to zero for \(\boldsymbol{\mu}_{n}\) and \(\boldsymbol{\Sigma}_{n} = \boldsymbol{I}\) for \(n=1,\,\ldots,\,N\). Kaiming initialization~\cite{He2015} was used to initialize the remaining parameters of all evaluated models. Training was conducted using the AdamW optimizer~\cite{Loshchilov2019} with batch-size~\(256\) and a learning rate of \(0.05\). The learning rate was varied by adopting the scheduling scheme from~\cite{Vaswani2017}. Separate models were trained for \(200\) epochs in each cross-validation fold. The models yielding the best validation loss were chosen for performance evaluation on the test sets.

\section{Results and discussion}
\label{sec:results}
\begin{figure}[t]
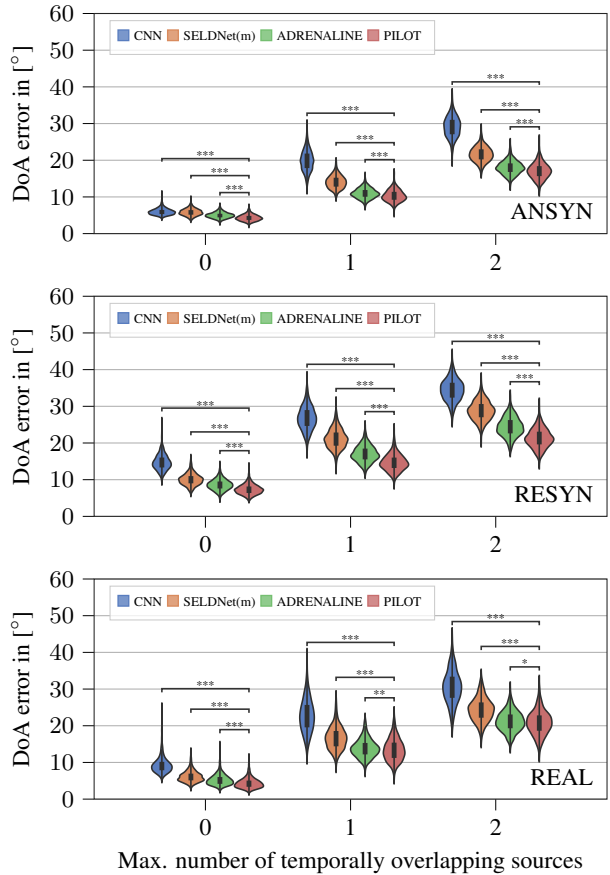

	\centering
	\input{figs/doa_error_ansim}%
	\input{figs/doa_error_resim}%
	\input{figs/doa_error_real}%
	\vspace*{-0.1cm}
\caption{Grouped violin-plots of \gls{DoA} errors obtained over all cross-validation folds for the ANSYN, RESYN and REAL datasets. Three asterisks (\({\ast\ast}\ast\)) indicate a statistically significant difference with \(p < 0.001\), two asterisks (\({\ast\ast}\)) correspond to \(p < 0.01\) and one asterisk (\({\ast}\)) denotes \(p < 0.05\).}
	\label{fig:doa_errors}
		\vspace*{-0.1cm}
\end{figure}

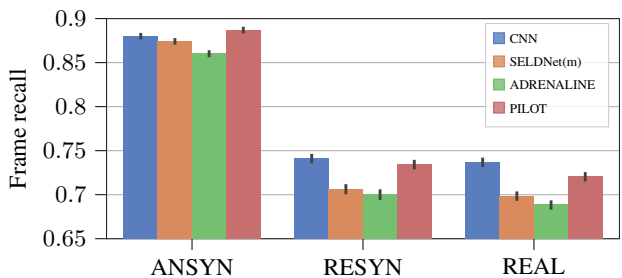
\begin{figure}[h!]
	\centering
\begin{tikzpicture}

\definecolor{color0}{rgb}{0.349019607843137,0.490196078431372,0.749019607843137}
\definecolor{color1}{rgb}{0.852941176470588,0.544117647058824,0.370588235294118}
\definecolor{color2}{rgb}{0.46078431372549,0.749019607843137,0.443137254901961}
\definecolor{color3}{rgb}{0.780882352941177,0.430882352941176,0.430882352941176}

\begin{axis}[
axis line style={white!15!black},
legend cell align={left},
legend style={fill opacity=0.8, draw opacity=1, text opacity=1, draw=white!80!black},
tick align=outside,
tick pos=left,
x grid style={white!69.0196078431373!black},
xlabel={},
xmin=-0.5, xmax=2.5,
xtick style={color=black},
xtick={0,1,2},
xticklabels={ANSYN,RESYN,REAL},
y grid style={white!69.0196078431373!black},
ylabel={Frame recall},
ymajorgrids,
ymin=0.65, ymax=0.9,
ytick style={color=black},
ytick={0.65, 0.70, 0.75, 0.80, 0.85, 0.90},
height=4.5cm,
width=0.49\textwidth
]
\draw[draw=none,fill=color0] (axis cs:-0.4,0) rectangle (axis cs:-0.2,0.879952839588518);
\addlegendimage{color0, only marks, mark=square*, line width=0.5pt,yshift=0.03cm};
\addlegendentry{\tiny CNN}

\draw[draw=none,fill=color0] (axis cs:0.6,0) rectangle (axis cs:0.8,0.741366666467259);
\draw[draw=none,fill=color0] (axis cs:1.6,0) rectangle (axis cs:1.8,0.736848395190852);
\draw[draw=none,fill=color1] (axis cs:-0.2,0) rectangle (axis cs:0,0.873927654622963);
\addlegendimage{color1, only marks, mark=square*, line width=0.5pt,yshift=0.03cm};
\addlegendentry{\tiny SELDNet(m)}

\draw[draw=none,fill=color1] (axis cs:0.8,0) rectangle (axis cs:1,0.706397777806407);
\draw[draw=none,fill=color1] (axis cs:1.8,0) rectangle (axis cs:2,0.698539259600111);
\draw[draw=none,fill=color2] (axis cs:2.77555756156289e-17,0) rectangle (axis cs:0.2,0.859997283971815);
\addlegendimage{color2, only marks, mark=square*, line width=0.5pt,yshift=0.03cm};
\addlegendentry{\tiny ADRENALINE}

\draw[draw=none,fill=color2] (axis cs:1,0) rectangle (axis cs:1.2,0.699984197782556);
\draw[draw=none,fill=color2] (axis cs:2,0) rectangle (axis cs:2.2,0.688503209947852);
\draw[draw=none,fill=color3] (axis cs:0.2,0) rectangle (axis cs:0.4,0.886940000187111);
\addlegendimage{color3, only marks, mark=square*, line width=0.5pt,yshift=0.03cm};
\addlegendentry{\tiny PILOT}

\draw[draw=none,fill=color3] (axis cs:1.2,0) rectangle (axis cs:1.4,0.734338272114741);
\draw[draw=none,fill=color3] (axis cs:2.2,0) rectangle (axis cs:2.4,0.720338271535889);
\addplot [line width=1.08pt, white!26!black, forget plot]
table {%
	-0.3 0.87630833961693
	-0.3 0.883701783811329
};
\addplot [line width=1.08pt, white!26!black, forget plot]
table {%
	0.7 0.735579685192748
	0.7 0.746403339313346
};
\addplot [line width=1.08pt, white!26!black, forget plot]
table {%
	1.7 0.731450697625664
	1.7 0.742146141620996
};
\addplot [line width=1.08pt, white!26!black, forget plot]
table {%
	-0.1 0.870395981930003
	-0.1 0.877786494553197
};
\addplot [line width=1.08pt, white!26!black, forget plot]
table {%
	0.9 0.700421907670881
	0.9 0.711971938512882
};
\addplot [line width=1.08pt, white!26!black, forget plot]
table {%
	1.9 0.69300811784848
	1.9 0.7036827291865
};
\addplot [line width=1.08pt, white!26!black, forget plot]
table {%
	0.1 0.856213889116146
	0.1 0.864121444147715
};
\addplot [line width=1.08pt, white!26!black, forget plot]
table {%
	1.1 0.694100574232019
	1.1 0.706098963746986
};
\addplot [line width=1.08pt, white!26!black, forget plot]
table {%
	2.1 0.683057703949479
	2.1 0.693604826873689
};
\addplot [line width=1.08pt, white!26!black, forget plot]
table {%
	0.3 0.883133333363066
	0.3 0.890743605156517
};
\addplot [line width=1.08pt, white!26!black, forget plot]
table {%
	1.3 0.728829062262799
	1.3 0.739628642161396
};
\addplot [line width=1.08pt, white!26!black, forget plot]
table {%
	2.3 0.714950296259297
	2.3 0.725621857935882
};
\end{axis}

\end{tikzpicture}%
	\vspace*{-0.4cm}
	\caption{Frame recall of all investigated methods with standard deviations (denoted by the black bars), averaged over all cross-validation folds.}
	\label{fig:frame_recall}
		\vspace*{-0.2cm}
\end{figure}

The experimental results are depicted in Figs.~\ref{fig:doa_errors} and~\ref{fig:frame_recall}. An initial application of the Shapiro-Wilk test~\cite{Shapiro1965} on the resulting \gls{DoA} errors rejected the null-hypothesis of normally distributed samples. Hence, the one-sided Mann-Whitney-\(U\) test~\cite{Mann1947} was used to assess statistically significant differences in \gls{DoA} error between the proposed \gls{PILOT} framework and the three baselines.

The results indicate that the proposed \gls{PILOT} framework outperforms all three baseline methods in terms of \gls{DoA} error with statistically significant differences in all evaluation scenarios. Notably, all methods that incorporate temporal context via either \glspl{RNN} or transformers yield improved performance over the \gls{CNN} baseline. However, it has to be taken into account here that the \gls{CNN} baseline has significantly fewer parameters than the other models. A direct comparison between ADRENALINE and \gls{PILOT} reveals that, even though both models have nearly the same capacity in terms of trainable parameters, the transformer-based \gls{PILOT} framework consistently outperforms the \gls{RNN}-based baseline model. This finding provides empirical evidence that transformers can be successfully applied to \gls{SEL} tasks on the datasets evaluated in this study. Additionally, \gls{PILOT} comes up with estimation uncertainties, which none of the other frameworks is able to provide, cf. Fig.~\ref{fig:estimation_comparison}.

In terms of frame recall, the proposed \gls{PILOT} model yields significantly better results than the modified SELDNet and ADRENALINE baselines on the evaluated datasets. All frameworks show a similar drop in performance for estimating the number of sound events in reverberant environments. This indicates that reverberation entails a challenge for distinguishing between concurrent sound events, even though temporal context is explicitly utilized by the proposed \gls{PILOT} system.

\section{Conclusion and outlook}
\label{sec:conclusion}
In summary, the results presented in this work provide a first promising perspective on how transformer-based models can be exploited for \gls{SEL} tasks. The proposed \gls{PILOT} framework shows that such models contribute to beneficial performance in \gls{SEL} tasks compared to \gls{RNN}-based systems. Moreover, a differentiable probabilistic localization output stage based on a linear Gaussian system was introduced that provides a means to estimate uncertainties for deep learning-based sound event localization. Future work might focus on incorporating these uncertainties into suitable applications and evaluating their benefits to tasks related to, e.g., decision making.


\bibliographystyle{IEEEtran}
\bibliography{refs}

\end{document}